# Quantifying truth and authenticity in AI-assisted candidate evaluation: A multi-domain pilot analysis

E. Lee [1, *], N. Worley [1], K. Takatsuji [1]

[1] *Altera Strategy Foundry Inc., Mountain View, CA, USA*

This paper presents a retrospective analysis of anonymized candidate-evaluation data collected during pilot hiring campaigns conducted through *AlteraSF*, an AI-native résumé-verification platform. The system evaluates résumé claims, generates context-sensitive verification questions, and measures performance along quantitative axes of factual validity and job fit, complemented by qualitative integrity detection. Across six job families and 1,700 applications, the platform achieved a 90-95% reduction in screening time and detected measurable linguistic patterns consistent with AI-assisted or copied responses. The analysis demonstrates that candidate truthfulness can be assessed not only through factual accuracy but also through patterns of linguistic authenticity. The results suggest that a multi-dimensional verification framework can improve both hiring efficiency and trust in AI-mediated evaluation systems.

## 1. Introduction

The modern hiring landscape faces a growing challenge of résumé inflation and AI-assisted deception. Research consistently shows that between 30% and 70% of applicants exaggerate or falsify their qualifications. A 2025 survey by FlexJobs found that 33% of respondents admitted to lying on résumés or cover letters [1]. ResumeLab's 2023 study reported that 70% of applicants had done so at least once, and 37% said they had done so regularly [3]. Glassdoor and Forbes have documented similar findings, with hiring managers increasingly rejecting candidates based on discovered discrepancies [2,4].

Generative AI tools now enable job seekers to produce highly fluent yet unverifiable content in seconds. This widening gap between surface fluency and factual truth undermines recruiter confidence and increases time spent validating claims manually. Traditional applicant tracking systems (ATS) such as Ashby, Greenhouse, Lever, and Workday were built for workflow management and candidate collaboration, not factual verification. They remain limited to structured scoring and pipeline analytics without mechanisms to evaluate whether résumé claims are true.

*AlteraSF* was developed to close this gap. The system extracts factual claims from résumés, generates dynamic verification questions, and scores responses along two quantitative axes: (1) Claim Validity (truth likelihood) and (2) Job Fit (contextual alignment). A third qualitative dimension, Integrity Signals, was introduced to detect linguistic or behavioral anomalies such as repetitive phrasing, copy-paste behavior, or suspected AI assistance, but these indicators are flagged for review rather than scored numerically.

* Correspondence to eldred@alterasf.com

This study analyzes anonymized data from live hiring campaigns using *AlteraSF* to quantify its impact on accuracy, speed, and authenticity detection across multiple job families. The findings demonstrate how automated verification can reduce manual review by over 90%, surface integrity anomalies, and establish measurable transparency in AI-assisted hiring.

## 2. Data Sources and Ethics

All data were collected during normal use of the *AlteraSF* platform. Candidates applied through standard postings and answered automatically generated verification questions as part of routine hiring workflows. No experimental interventions were introduced. After campaigns concluded, data were anonymized and aggregated for analysis.

No personally identifiable information (PII) was retained or reviewed. The study does not constitute human-subjects research under institutional review definitions, as no individual behavior was manipulated or evaluated beyond standard hiring operations. Analyses were conducted solely for system validation and improvement. All examples have been generalized for privacy.

## 3. System Overview

The *AlteraSF* architecture uses transformer-based natural language models for entity recognition and claim clustering. Each claim is embedded contextually and matched against role requirements to produce a Claim Validity score. A Cross-Validation Matrix then maps Claim Validity against Job Fit. Version 1.1 adds integrity detection by analyzing entropy variance, token repetition, and syntactic regularity. These results are presented as flags rather than scores, enabling recruiters to inspect specific anomalies.

| Version | Core Capabilities | Description |
|---|---|---|
| **v0.9** (Pre-Analytics) | Claim Validity + Job Relevancy | Baseline version validating résumé statements against job descriptions without analytics or integrity detection. |
| **v1.0** (Analytics Layer) | Cross-Validation Matrix + ROI Analytics | Introduced analytics to visualize completion rates, time savings, and "diamond" yield (top candidates). |
| **v1.1** (Integrity Layer) | Linguistic & Behavioral Integrity Detection | Added qualitative detection of AI-assisted phrasing, copy-paste repetition, and entropy collapse. |

Table 1. Summary of *AlteraSF*'s version evolution: v0.9 focused on factual verification, v1.0 introduced measurable analytics (completion rate, time savings, ROI), and v1.1 added integrity detection to flag possible AI or copied content. Claim Validity and Job Fit are quantitative scores, while Integrity Signals are qualitative flags.

## 4. Methodology

Each pilot followed a standardized six-stage evaluation workflow:

1. **Job Posting**: Live roles were listed through *AlteraSF* for open application.

2. **Claim Extraction**: The model segmented résumés into factual statements (skills, metrics, outcomes) using dependency parsing and named-entity recognition.

3. **Question Generation**: Dynamic questions were generated from extracted claim. The model produces unique phrasing for each candidate, ensuring non-deterministic prompts.

4. **Scoring**: Candidate responses were analyzed for factual coherence, with Claim Validity and Job Fit each normalized to a 0-5 scale. Candidates scoring >4 in Claim Validity and exactly 5 in Job Fit were considered cross-validated and designated as "Diamonds," representing high-confidence applicants whose résumés and verification responses demonstrated both factual reliability and complete alignment with the target role.

5. **Analytics Capture**: From v1.0 onward, the platform recorded completion rates, time efficiency, and ROI impact (cost saved).

6. **Integrity Detection**: Beginning with v1.1, the system tracked linguistic entropy, token burst variance, and repetitive phrasing patterns to identify possible AI assistance.

All datasets were processed after the conclusion of hiring cycles. Individual applicant data were never shared or interpreted outside aggregated metrics.

## 5. Results by Domain

The following subsections summarize outcomes from six pilot hiring campaigns analyzed between versions v0.9 and v1.1. Each dataset highlights the evolution of *AlteraSF*'s performance in claim verification, contextual alignment, and qualitative integrity detection. As aforementioned in Section 4, "Diamonds Found" refers to candidates achieving both Claim Validity > 4 and Job Fit = 5.

*5.1 International Sales Role (v0.9 ➔ v1.0)*

| Metric | Value |
|---|---|
| Applicants | 39 |
| Completion Rate | 25.6% |
| Diamonds Found | 0 |
| Claim Validity Mean | 2.7 ± 0.68 |
| Job Fit Mean | 3.2 ± 0.67 |
| Time Efficiency | 6.5hr saved |
| ROI Impact | $325 |
| Screening Speed | 40× faster |

Table 2. Baseline multilingual sales pilot under v0.9 ➔ v1.0 validated question randomness and multilingual robustness despite no cross-validated "diamond" candidates.

This pilot evaluated a remote commission-based sales position recruiting contractors in Southeast Asia and Eastern Europe. Résumés were concise and multilingual. Claim Validity scores were low due to limited quantifiable information, but contextual relevance remained strong. Several candidates reapplied using identical résumés to test question consistency and found that each attempt generated different prompts, confirming stochastic question generation. These results provided early empirical validation of non-deterministic claim verification and informed multilingual normalization.

*5.2 Domestic Sales Role (v0.9 → v1.0)*

| Metric | Value |
|---|---|
| Applicants | 8 |
| Completion Rate | 25.0% |
| Diamonds Found | 0 |
| Claim Validity Mean | 2.4 ± 0.60 |
| Job Fit Mean | 3.2 ± 0.40 |
| Time Efficiency | 1.3hr saved |
| ROI Impact | $67 |
| Screening Speed | 28× faster |

Table 3. Small-sample domestic dataset showing authentic human language variability and establishing the entropy baseline later used for calibration in v1.1.

This domestic sales role targeted US-based contractors in live marketing and events. Résumés were informal and story-driven, with high linguistic variability. The model detected mid-range fit but low measurable validity, reflecting unstructured writing styles. Despite this, the analytics layer identified the most relevant candidates while maintaining high authenticity. This dataset became an early benchmark for entropy-based calibration in v1.1.

*5.3 Creative Design Role (v1.1)*

| Metric | Value |
|---|---|
| Applicants | 74 |
| Completion Rate | 10.8% |
| Diamonds Found | 1 |
| Claim Validity Mean | 3.4 ± 0.80 |
| Job Fit Mean | 4.5 ± 0.60 |
| Time Efficiency | 12.3hr saved |
| ROI Impact | $613 |
| Screening Speed | 148× faster |

Table 4. Creative-domain pilot under v1.1 showed high contextual alignment and first recorded integrity flag for repetition, establishing behavioral baselines for linguistic-entropy tracking.

The creative-design role tested *AlteraSF* v1.1, which incorporated analytics and integrity detection. Applicants emphasized collaboration, visual systems, and process storytelling rather than numerical metrics. One high-performing candidate achieved "diamond" status but displayed repeated phrasing later flagged as an integrity anomaly.

*5.4 Marketing Role (v1.1)*

| Metric | Value |
|---|---|
| Applicants | 107 |
| Completion Rate | 13.1% |
| Diamonds Found | 0 |
| Claim Validity Mean | 3.1 ± 0.50 |
| Job Fit Mean | 3.8 ± 0.70 |
| Time Efficiency | 17.8hr saved |
| ROI Impact | $892 |
| Screening Speed | 148× faster |

Table 5. Marketing-communication dataset under v1.1 displayed the lowest flag rate (~4%), confirming robustness of integrity detection in narrative domains.

Marketing applicants produced narrative-heavy, expressive résumés centered on brand messaging and campaign impact. The v1.1 framework maintained accurate extraction and filtering even without strong quantitative cues. Only ~4% of submissions triggered low-confidence integrity flags, indicating generally authentic human writing.

*5.5 Software Role (v1.1)*

| Metric | Associate | Engineer |
|---|---|---|
| Applicants | 566 | 819 |
| Completion Rate | 37.3% | 22.6% |
| Diamonds Found | 12 | 18 |
| Claim Validity Mean | 3.3 ± 0.80 | 3.1 ± 0.75 |
| Job Fit Mean | 4.3 ± 1.00 | 4.8 ± 0.80 |
| Time Efficiency | 93.3hr saved | 135.0hr saved |
| ROI Impact | $4,667 | $6,750 |
| Screening Speed | 94.3× faster | 91× faster |

Table 6. Software datasets under v1.1 demonstrated scalability and clear separation between factual accuracy and linguistic originality.

The software-domain pilots represented the largest datasets and introduced full-scale integrity analytics. Among 1,385 applicants, 30 diamonds met the dual thresholds. Approximately 12% of total submissions and ~60% of diamonds exhibited low-entropy, repetitive syntax typical of AI-assisted text. Despite this, factual coherence remained strong, confirming that validity and authenticity are orthogonal variables.

*5.6 Hardware Role (v1.1)*

| Metric | Value |
|---|---|
| Applicants | 13 |
| Completion Rate | 53.8% |
| Diamonds Found | 1 |
| Claim Validity Mean | 3.4 ± 0.90 |
| Job Fit Mean | 4.3 ± 0.70 |
| Time Efficiency | 2.1hr saved |
| ROI Impact | $105 |
| Screening Speed | 26× faster |

Table 7. Hardware engineering dataset under v1.1 confirmed cross-domain applicability of integrity detection to scientific and R&D contexts.

The hardware-domain pilot assessed complex, technical résumés in materials and device engineering. Candidates typically held advanced degrees and used specialized terminology. The system parsed such content accurately and flagged one candidate for syntactic uniformity suggestive of AI refinement.

## 6. Integrity Signal Analysis

| Domain | Version | Cheating Signal Rate | Observation |
|---|---|---|---|
| Creative Design | v1.1 | High (1/1 Diamond) | Repetition, copy-paste anomaly |
| Marketing | v1.1 | Low (~4%) | Authentic human narrative style |
| Software | v1.1 | High among top tier (~60%) | AI-style regularity patterns |
| Sales (Int'l + Domestic) | v1.0 | Minimal | Human linguistic diversity |
| Hardware | v1.1 | Moderate | Partial AI assistance flag (1 candidate) |

Table 8. Integrity-flag summary across all roles. Rates rise with domain technicality, suggesting that greater structure invites generative assistance or templating behavior.

Integrity detection in v1.1 identified three primary anomaly categories: textual repetition, entropy collapse, and response homogeneity. Flagged content typically indicated AI-assisted drafting rather than intentional falsification. The trend correlates with domain technicality: Integrity flags appeared most frequently in structured, code-adjacent writing and least in narrative or interpersonal fields.

## 7. ROI and Efficiency Analysis

| Parameter | Symbol | Value |
|---|---|---|
| Recruiter hourly rate | $R$ | $50 |
| Traditional review time | $T$ | 10 min/applicant |
| AI-assisted review time | $t$ | 2 min/applicant |
| Total applicants | $N$ | Total number of applications received |
| Reviewed candidates | $n$ | Subset of applications reviewed (cross-validated "Diamonds") |

Table 9. Constants used for time efficiency, screening speed improvement, and ROI impact estimation. Calculations assume recruiter rate = $50/hr, 10 → 2-minute review reduction, and identical applicant volume N for domain comparability.

Using these constants, time savings and cost efficiency were estimated as:

$$\text{Time Saved} = (N \times T) - (n \times t)$$

$$\text{Cost Saved (ROI Impact)} = R \frac{(N \times T) - (n \times t)}{60}$$

$$\text{Speed Improvement} = \frac{(N \times T)}{(n \times t)}$$

Average screening time decreased by 90-95 %, translating to 28-150× faster processing across pilots. Even conservative estimates yielded recruiter cost reductions between $50 and several thousand dollars per role, depending on applicant volume.

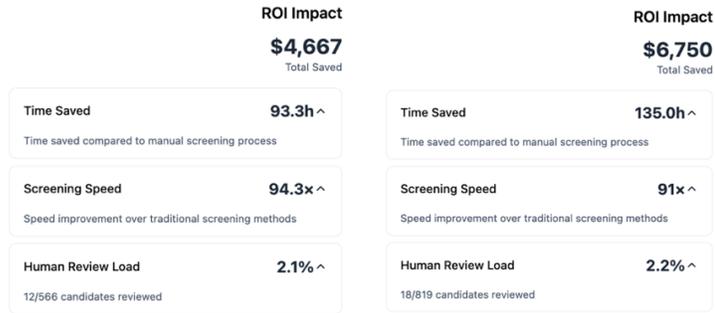

Figure 1. ROI impact visualization of software roles.

Because *AlteraSF* narrows review scope to cross-validated "diamond" candidates (N → n), savings compound linearly with scale.

## 8. Discussion

The multi-domain results confirm that combining quantitative claim verification with qualitative authenticity detection significantly improves screening efficiency without compromising interpretability. While Claim Validity × Job Fit scores drive ranking, qualitative integrity indicators contextualize those results, showing how candidates articulate expertise, not just what they claim.

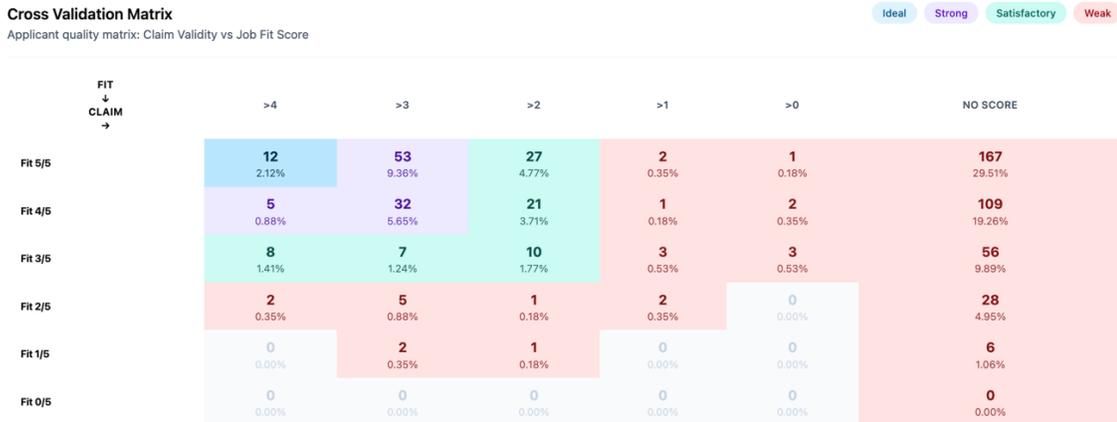

Figure 2. Example of Cross-Validation Matrix (Claim Validity × Job Fit).

Across domains, technical and engineering candidates displayed the highest rate of integrity flags, likely reflecting widespread use of generative drafting tools. Creative and marketing applicants produced more varied linguistic structures, yielding fewer anomalies. This demonstrates that factual precision and linguistic authenticity behave as independent but complementary variables.

From a governance perspective, *AlteraSF* introduces transparency often missing in AI-based hiring. Unlike opaque ranking systems, the model exposes both the claim-level evidence and the reasoning paths (i.e. specific question-answer validation). This transparency enables auditors to understand why candidates rank highly, critical for compliance under emerging algorithmic-accountability regulations.

## 9. Future Work and Limitations

Current limitations primarily concern sample size and domain diversity, not functionality. While *AlteraSF* successfully processed multilingual résumés in the international sales pilot, demonstrating cross-language consistency and unique question generation, the breadth of linguistic benchmarking remains limited. Future expansions will extend testing across additional language families and refine entropy calibration for non-Latin scripts.

False positives occasionally arise when legitimate technical conciseness mimics generative uniformity. To mitigate this, entropy thresholds will be dynamically adjusted by domain (i.e. differentiating between the structured brevity of engineering roles and the narrative style of design or policy roles).

Ongoing research also explores temporal signature modeling, analyzing response cadence and hesitation patterns to distinguish genuine cognitive latency from automated consistency.

Finally, partnerships with enterprise HR teams and academic institutions need to be undergone to formalize "truth integrity" metrics for AI-mediated evaluation, establishing empirical standards for verifiable trust in hiring.

## 10. Positioning Relative to ATS Platforms

Modern ATS primarily manage logistics: Scheduling, structured scorecards, and compliance.

*Greenhouse* emphasizes structured interviewing [6]; *Lever* focuses on pipeline analytics [7]; *Workday Recruiting* integrates "HiredScore AI" for large-scale matching [8]; and *Ashby* recently added Fraudulent Candidate Detection (Sept. 2025) [5].

*AlteraSF* complements these systems rather than replacing them. It provides claim-level verification, non-deterministic question generation, and qualitative integrity detection, operating either as a stand-alone lightweight ATS for SMBs and founders or as an embedded analytics layer within enterprise ATS ecosystems.

| Capability | AlteraSF | Ashby | Greenhouse | Lever | Workday Recruiting |
|---|---|---|---|---|---|
| **Claim-level factual verification** | Yes | Not stated | Not stated | Not stated | Not stated |
| **Dynamic, non-deterministic claim Q&A** | Yes | Not stated | Not stated | Not stated | Not stated |
| **Integrity analysis** | Yes (claim-level detection and logging of linguistic anomalies) | Fraudulent-candidate detection | Not stated | Not stated | Not stated |
| **Structured hiring / scorecards** | Via analytics overlays | Core | Core | Core | Core |
| **End-to-end ATS workflow** | Integrates alongside enterprise ATS; also functions as lightweight ATS for founders and SMBs (AI co-pilot) | Yes | Yes | Yes | Yes |
| **Recent AI features of note** | Integrity detection linked to verified claims | Fraud-detection automation | AI-enhanced structured hiring | Pipeline analytics | HiredScore AI matching |

Table 10. Comparative positioning of *AlteraSF* relative to leading ATS vendors. Unlike legacy ATS platforms optimized for logistics, *AlteraSF* performs factual verification and authenticity analysis, serving both enterprise integrations and lightweight use cases.

The growing scrutiny around algorithmic bias, exemplified by Workday's 2024 AI-bias litigation [9], underscores the need for transparent, auditable AI. By separating verification from ranking, *AlteraSF* offers explainable outputs that strengthen fairness and compliance.

## 11. Conclusion

Through multi-domain pilot analyses, *AlteraSF* demonstrated that automated claim-level verification paired with qualitative integrity detection can reduce recruiter workload by over 90%, increase screening speed 28-150×, and surface truthful, high-fit candidates with measurable transparency.

Where traditional ATS platforms track processes, *AlteraSF* verifies substance, confirming what candidates claim, not just how they apply. This structural difference establishes a new benchmark for evidence-based hiring. By linking factual accuracy with authenticity cues, *AlteraSF* directly addresses the credibility challenges posed by generative AI résumés. It integrates seamlessly with enterprise ATS systems or functions as a lightweight ATS for smaller teams, making rigorous verification accessible across company scales.

Future iterations will expand multilingual calibration, refine entropy thresholds, and formalize explainable-AI standards for hiring.

Ultimately, *AlteraSF* is built not merely to manage applicants but to measure truth and operationalize trust, positioning it as the foundational verification layer for the next generation of AI-driven recruiting systems.

## Acknowledgements

The authors thank now-offboarded design technologist (unnamed) for design and user-experience support during v1.1 development, and the early pilot collaborators for providing anonymized recruiting data that enabled empirical validation of the system.

# Appendix A. Supplemental Data

## A.1 Dataset Summary

| Domain | Version | Applicants | Diamonds Found | Completion Rate (%) | Integrity Flag Rate (%) |
|---|---|---|---|---|---|
| International Sales | v0.9 → v1.0 | 39 | 0 | 25.6 | 0 |
| Domestic Sales | v0.9 → v1.0 | 8 | 0 | 25.0 | 0 |
| Creative Design | v1.1 | 74 | 1 | 10.8 | 6 |
| Marketing | v1.1 | 107 | 0 | 13.1 | 4 |
| Software (Associate + Engineer) | v1.1 | 1,385 | 30 | 30.1 (avg) | 12 total / ~60 of Diamonds |
| Hardware Engineering | v1.1 | 13 | 1 | 53.8 | 8 |

Table A1. Aggregate summary of all pilot datasets. "Integrity Flag Rate" denotes proportion of applicants with detected linguistic anomalies such as repetition or entropy collapse. "Diamonds Found" refers to candidates meeting Claim Validity >4 and Job Fit = 5 thresholds.

## A.2 ROI Sensitivity Analysis

| Applicant Volume | Traditional Time (hr) | AI-Assisted Time (hr) | Time Saved (hr) | Cost Saved (@ $50/hr) |
|---|---|---|---|---|
| 50 | 8.3 | 1.7 | 6.6 | $330 |
| 250 | 41.7 | 8.3 | 33.4 | $1,670 |
| 500 | 83.3 | 16.7 | 66.6 | $3,330 |
| 1,000 | 166.7 | 33.3 | 133.4 | $6,670 |

Table A2. ROI sensitivity analysis using recruiter rate = $50/hr, 10 min → 2 min screening time reduction per applicant. Cost savings scale linearly with applicant volume; large technical roles yield several-thousand-dollar reductions per posting.

## A.3 Glossary of Key Terms

| Term | Definition |
|---|---|
| Claim Validity | Quantitative score (0-5) estimating factual consistency between résumé statements and candidate responses |
| Job Fit | Quantitative score (0-5) measuring contextual alignment with role requirements |
| Diamond Candidate | Applicant with Claim Validity >4 and Job Fit = 5; prioritized for recruiter review |
| Integrity Signal | Qualitative flag indicating linguistic or behavioral anomalies (i.e. repetition, entropy collapse, or generative regularity) |
| Entropy Collapse | Reduction in lexical or syntactic diversity characteristic of AI-generated text |
| Token Burst Variance | Variation in sentence length and structure across sequential text bursts; low variance indicates automated output |